# Generating Empirical Core Size Distributions of Hedonic Games using a Monte Carlo Method

Andrew J. Collins*, Sheida Etemadidavan, and Wael Khallouli

Old Dominion University

Engineering Management & Systems Engineering

Norfolk, VA 23529, USA

ajcollin@odu.edu, setemadi@odu.edu, wkhallou@odu.edu

*Corresponding Author

*ORCID 0000-0002-8012-2272

# Abstract

Data analytics allows an analyst to gain insight into underlying populations through the use of various computational approaches, including Monte Carlo methods. This paper discusses an approach to apply Monte Carlo methods to hedonic games. Hedonic games have gain popularity over the last two decades leading to several research articles that are concerned with the necessary, sufficient, or both conditions of the existence of a core partition. Researchers have used analytical methods for this work. We propose that using a numerical approach will give insights that might not be available through current analytical methods. In this paper, we describe an approach to representing hedonic games, with strict preferences, in a matrix form that can easily be generated; that is, a hedonic game with randomly generated preferences for each player. Using this generative approach, we were able to create and solve, i.e., find any core partitions, of millions of hedonic games. Our Monte Carlo experiment generated games with up to thirteen players. The results discuss the distribution form of the core size of the games of a given number of players. We also discuss computational considerations. Our numerical study of hedonic games gives insight into the underlying properties of hedonic games.

# 1 Introduction

The age of big data and data analytics is upon us, so it seems reasonable to ask how this emerging discipline could apply to game theory. Data analytics is concerned with the extraction, storage, manipulation, analysis, and interpretation of empirical data. It is a new field that sits at the intersection of the fields of computer science, operations research, and statistics. Monte Carlo methods are used in data analytics to generate and interpret data (Lustig, Dietrich, Johnson, & Dziekan, 2010). In this paper, we apply a Monte Carlo method to generate millions of hedonic games and find their core partition sets, which we call the core. By generating this vast number of games, and their stable solutions, allows us to create empirical distributions of the outcomes. These empirical distributions give insight into the underlying properties of hedonic games, which might not be obtainable through analytical means. By understanding these underlying qualities, researchers might better understand hedonic games and, more importantly, a better understanding of how they may be applied to real-world problems.

Data analytics, data analysis, or plainly, analytics is an emerging field used to gain understanding from data. Though both data analytics and analytical methods derive their names from the word analysis, it should be noted that the two methodologies are very different (Park, 2017). Data analytics is concerned with the holistic study of empirical data, whereas analytical methods are concerned with logical deduction of the generalized phenomenon using mathematical proofs (note that analytical methods are also different from mathematical analysis, which focuses on the study of limits and advanced calculus). Analytical methods have dominated the research techniques used in game theory (though behavioral game theorists have used human experiments for many years, see Axelrod (1984) or Güth, Schmittberger, and Schwarze (1982) for classic examples). We are proposing that numerical data analysis is a useful tool for game theory researchers, and this paper gives a demonstration of this potential application.

The game type considered in our study is a special type of cooperative game theory known as hedonic games (Banerjee, Konishi, & Sönmez, 2001; Bogomolnaia & Jackson, 2002). Hedonic games are a form of non-transferable utility (NTU) game where an individual's payoff is purely determined by the membership makeup of their current coalition. A hedonic game is usually represented as a collection of preference relations; each represents a single players' preferences over the collection of possible coalitions that they could be a member.

The reason that we used hedonic games for our study is that it removes one of the problems of cooperative game theory. Technically, finding the core of a cooperative game is a simultaneous two-part problem; first, you need to determine which coalitions will form and, secondly, you need to determine how the payoff distributes within each coalition (this payoff distribution is called an imputation). Hedonic games remove the second problem because the utility cannot be transferred between players, i.e., the preferences for a given player in a given coalition is fixed[1]. The first problem can be removed by assuming that the game has the super-additivity property (because this will result in the formation of the grand coalition); thus, the focus of super-additive games is determining the imputation of the game. Hedonic games are not necessarily super-additive and, as such, suffer from the first problem, that is, determining which coalitions will form.

Since hedonic games focus on coalition formation, not payoff allocation, means that they are also called coalition formation games (İnal, 2019). This is not the same as canonical coalition games, which tend to assume the super-additivity property (Saad, Han, Debbah, Hjorungnes, & Basar, 2009). Finding the core partition in a super-additive hedonic game is trivial, i.e., its grand coalition. In this paper, we focus on randomly generated hedonic games that may or may not be super-additive. The core partition for these randomly generated games is found through an exhaustive search.

---

[1] Technically, for an NTU game, the payoff to a player, in a given coalition, does not have to be fixed as a coalition might have multiple choices of payoffs available. In a hedonic game, the preferences, for a given coalition, is fixed for all players.

Cooperative game theory is also known as n-person game theory (Shapley, 1953); however, the name n-person is misleading because it implies that it is possible to consider many players in the game when, in reality, finding the core of, for instance, a score of players is computationally intractable. As such, our numerical analysis of hedonic games is limited to about thirteen players. Although this number seems small, the sear number of coalitions and coalition structures that must be checked is staggering, which we will discuss later.

Given the computational difficulties in finding the core partition of large hedonic games it is unsurprising that the literature on hedonic games has focused on finding analytical solutions, especially; necessary, sufficient, or both conditions for core stability (Banerjee et al., 2001; Bogomolnaia & Jackson, 2002; Dimitrov & Sung, 2007; Iehlé, 2007; İnal, 2019). However, though analytical solutions provide insight nature of hedonic games, their conditions can be as difficult to find, in practice, as finding the core itself, e.g., I-balanced condition (Iehlé, 2007). Also, focusing on analytical results does not give a sense of what the general properties of real solutions to hedonic games are. As such, our research concern is a more simplistic understanding of the core of hedonic games through empirically generated results.

Another analytical approach used in the study of hedonic games is to restrict the form of the game to ensure core membership; this approach has been applied by some researchers (Alcalde & Romero-Medina, 2006; Apt & Witzel, 2009; Dimitrov, Borm, Hendrickx, & Sung,

2006; Hasan, Gorce, & Altman, 2020; Pápai, 2004). Others have created interesting variations of hedonic games to solve (Karakaya & Klaus, 2017). We are interested in knowing the general properties of hedonic games; that is, given a hedonic game, how frequently would expect the game not to be core stable? How many core partitions would you expect the game to have? In theory, the answers to these questions could be solved analytically. In practice, it is not obvious whether the mathematical tools required to solve these questions even exist yet.

There have been other attempts to understand the form of random hedonic games; for example, Pittel (2019) investigates, analytically, the properties of a randomly generated roommate game; they are able to conclude the probability of finding a stable matching reduces to zero as the number of players increases. We are interested in the general properties of hedonic games, not just a subset of hedonic game types, and we believe that to find general properties, we have to look beyond analytical approaches. This is especially true in a practical setting.

Solving a cooperative game in a practical setting, even for the simplistic case, can be quite challenging. As such, those that have used cooperative games, for real-world problems, have focused on a few players. For example, Salamanca (2016) only considers three players in their game, and Su (2019) limits the coalition (team) size to two. Recently, Toumasatos and Steinshamn (2018) managed to solve a practical five-player game of Northsea fishing competition. To make cooperative games, including hedonic games, useful for understanding problems involving more than a few players requires a shift from the analytical mindset.

We argue that given difficultly of analytically solving hedonic games, there is a demand for numerical approaches to gain insight into the hedonic games if they are going to be used in a practical setting. Numerical approaches can generate empirical distributions of properties of interest (of hedonic games), and these empirical distributions might give insight into the general properties of hedonic games. Given we are in the age of big data, these empirical distributions can be generated from millions of data points, making it highly unlikely that they differ greatly from any analytically derived distributions.

There are several numerical approaches available to us to investigate hedonic games, e.g., agent-based simulation, evolutionary game, artificial intelligence, algorithmic game theory or, Monte Carlo methods. We could approximate the solution mechanism in an agent-based simulation (Collins & Frydenlund, 2018; Szilagyi, 2007); however, such an approach does not guarantee a core partition will be found (Vernon-Bido & Collins, 2020). We could aggregate the players to create population games and solve using evolutionary games (Weibull, 1995), but it is not clear that the results of the approximated game would reflect that the original game considered. It has been shown that it is hard to know if anything has been learned using Artificial Intelligence approaches in situations involving more than two decision-makers (Collins, Sokolowski, & Banks, 2014). Algorithmic game theory is more focused on algorithm design than solving generic games. Thus, given the weakness of these other approaches, our approach was to use a Monte Carlo method to generate the core partition sets of millions of randomly generated hedonic. Since the actual core partitions are found for each game, if they exist, we ensure our results reflect the actual reality of hedonic games. Through the generation of millions of hedonic games and their core partitions (if they exist), we can create empirical distributions of certain properties of hedonic games, i.e., the distribution of the size of the core partition set.

## 1.1 Overview of paper

In this paper, we describe an approach for generating random hedonic games, and we use this approach to generate millions of hedonic games to find their core partition sets. The random generation of a hedonic game is done by uniquely representing each game in a matrix format. Using this numerical representation of a game enables us to find the core partition sets, of a given game, through a brute force algorithmic approach, i.e., each possible coalition is compared to each possible partition to see if the coalition blocks the partition. Due to the sheer numbers of partitions that need to be evaluated, a special algorithm was employed to ensure that the list of possible partitions was reviewed systematically. The technical details of these approaches will be provided in the Methods section of this paper.

The next section gives a brief overview of hedonic games and core stability; it also provides a brief overview of Monte Carlo methods. After the Methods section, the results from the Monte Carlos Experiment are described, and conclusions stated.

## 2 Background

Data analytics is a new field that has arisen due to the massive increase in available data that has occurred in modern times. As a new field, there is also no agreed-upon standard definition (Davenport & Harris, 2017; INFORMS, 2019; Lustig et al., 2010; National Academies of Sciences, 2017). The Institute for Operations Research and the Management Sciences (INFORMS) defines analytics as "the scientific process of transforming data into insight for making better decisions" (INFORMS, 2019). Davenport and Harris (2017) define analytics as "the extensive use of data, statistical and quantitative analysis, explanatory and predictive models, and fact-based management to drive decisions and actions." The National Academy of Sciences splits analytics into three subfields: descriptive analytics, predictive analytics, and prescriptive analytics (Lustig et al., 2010). They say that Monte Carlo methods are a predictive analytical technique and define predictive analytics as "the extensive use of data and mathematical techniques to uncover explanatory and predictive models of business performance representing the inherit relationship between data inputs and outputs/outcomes." It is our intention to uncover properties of hedonic games using Monte Carlo methods.

As mentioned previously, even the name "data analytics" is not used universally. The field is sometimes called data analysis, analytics, or data science. Technically, data science is the scientific implementation of data analytics and would require the scientific method to be followed, including hypothesis formation (National Academies of Sciences, 2017). The research presented here does not follow a hypothesis forming approach, and, as such, we claim it is data analytics. We believe that imposing the constraint of the scientific method on data analytics removes the flexibility required to implement data analytical techniques like data mining or machine learning[2].

A reader might find the lack of standardization within data analytics alarming; however, we would like to point out that other modern techniques suffer the same fate (Collins, Petty, Vernon-Bido, & Sherfey, 2015). We would also like to point out that the definitions relating to the cooperative game theory are not always been universal; for example, they were originally called n-person game theory (Rapoport, 1970; Shapley, 1953) and now sometimes called a coalitional game theory (however, this tends to refer to canonical coalition game theory or coalition formation game theory).

The remainder of this paper focuses on the method used, i.e., Monte Carlo, and the game type considered, i.e., hedonic games. Further discussion on data analytics can be found in Davenport and Harris (2017).

---

[2] https://www.kdnuggets.com/2016/11/machine-learning-vs-statistics.html

## 2.1 Hedonic Games

Since the turn of the century, hedonic games have gained interest in the academic community, because of their ability to model the grouping preferences of individuals (Aziz & Savani, 2016). Hedonic games are the generalization of matching games (Gale & Shapley, 1962; Roth & Sotomayor, 1992) like the stable roommate problem, and the marriage problem (Gale & Shapley, 1962). Both of these examples focus on players forming pairs, whereas the generic hedonic game problem considers coalitions of any size. Hedonic games have been applied to practical problems like organizing robotic swarms (Jang, Shin, & Tsourdos, 2018b).

There are several solution concepts suggested for hedonic games, including core stability, Nash stability, individually stability (Bogomolnaia & Jackson, 2002), and strong Nash stability (Karakaya, 2011). A hedonic game that is core stable is, in essence, the same as a game that has a non-empty core (İnal, 2019), and, due to the popularity of the core, we focus on finding core stable partitions in this paper. A core stable partition of players is one that is not blocked by any coalition. A coalition blocks a partition when all members of that coalition prefer that coalition over the coalitions they are currently assigned, i.e., they have an incentive to form the blocking coalition. Core partition is another name for a partition that is core stable (Banerjee et al., 2001). We will discuss core stability in more detail in the next section.

Hedonic games are a form of non-transferrable utility game where each player has a preference relation over subset for which they are a member. The concept of hedonic games was introduced by Dreze and Greenberg (1980) and mathematically formalized by Banerjee et al. (2001) and Bogomolnaia and Jackson (2002). The study of hedonic games is concerned with finding the stable coalition structure that forms among players. As such, hedonic games are also known as coalition formation games (Banerjee et al., 2001); this is not that same as canonical (coalition) games. Canonical coalition games are characteristic function games were the value of a coalition is dependent on all the current coalitions and not just the current coalition. Canonical coalition games tend to assume super-additive for ease of solving, as discussed above. Hedonic games do not assume super-additive, and the players' payoff is not dependent on what is occurring in other coalitions only who is a member of their coalition.

A coalition structure, π, is a collection of disjoint coalitions that covers all the players; it is equivalent to a complete partition of the set of players. Similarly, a coalition is equivalent to a subgroup of players. We assume that the game has a finite set of players, N. Whether stable coalition structures form is based on the preferences of the players. We represent these strict preferences as $\succ_i$ over the set of possible coalitions which a player is a member, $S(i) \subseteq 2^N$. Since the coalitions must be disjoint, no player can simultaneously be a member of multiple coalitions in the coalition structure. In this research, only strict preferences are allowed in the generated hedonic games; that is, $\forall s, t \in S(i)\ (s \succ_i t)\ \dot{\lor}\ (t \succ_i s)$.

### 2.1.1 Core stability

Core stability firstly proposed by Banerjee et al. (2001) and Bogomolnaia and Jackson (2002) as the core equivalent for hedonic games (İnal, 2019). A partition that is core stable is called a core partition (Banerjee et al., 2001). We have already defined core stability above, but here is the mathematical definition of a core partition:

$$\pi^* \in \Pi \text{ s.t. } \nexists S \in 2^N \text{ s.t } S \succ_i S_{\pi^*}(i)\ \forall i \in S$$

Where $\Pi$ is the collection of all possible partitions and $S_\pi(i)$ is the coalition in partition π that contains player *i*. Thus, a coalition partition is a partition where any subset of players does not have an incentive to deviate and form a new coalition. Note that for a given hedonic game, the existence of a core partition is not guaranteed; as the results later indicate, there is a significant chance of no core partition existing for a given random hedonic game.

Discussion on other solution concepts, like nash stability, individually stability, and contractual individually stability, can be found in Chalkiadakis, Elkind, and Wooldridge (2011).

### 2.1.2 Monte Carlo Methods

Monte Carlo (MC) methods are mathematical sampling approaches used to gain insight into theoretical properties through the generation of empirical distributions. They can also provide heuristics to solving optimization problems, i.e., simulated annealing (Mooney, 1997). A Monte Carlo simulation is the application of a Monte Carlo method to a simulation with the intention of understanding "what if" scenarios of some phenomenon of interest. The approach used in the paper is not a Monte Carlo simulation; however, MC simulation approaches have been applied to hedonic games; hence, the need to introduce them here.

To give a reader an understanding of MC methods, consider the simple example of trying to estimate the integral of a function, f, over the range of [0, 1]. Consider some $M \gg max\{f(x) : x \in [0,1]\}$ and two random variables $X \sim U(0,1)$ and $Y \sim U(0, M)$. The MC approach follows the following algorithm, for a set of N runs:

- Set c = 0, and n = 0
- Generate random variates x and y from X and Y respectfully
- If $f(x) \leq Y$ then $c \rightarrow c + 1$
- $n \rightarrow n + 1$
- If $n < N$ then go to step 2

The output of this algorithm can be used to estimate the integral of the f(x) as follows:

$$\int_0^1 f(x)\,dx \approx \frac{c.M}{n}$$

Thus, for functions where the integral is not known, the Monte Carlo method provides a means to find an estimate. If the Monte Carlo approach is applied multiple times for different upper bounds, this could provide insight into the functional form of the integral. It is a type of application that we apply, in this paper, to understand the distribution of the core size of hedonic games.

The roots of the Monte Carlo method stem back to Comte De Buffon needle experiments, to estimate the value of π, in 1733 (Cheng 2017) and Enrico Fermi, the "architect of the nuclear age," work in the 1930s (Cooper et al. 1989).

In 1945, Stanislaw Ulam and Nicholas Metropolis invented the computerized Monte Carlo method (Metropolis and Ulam 1949; Cooper et al. 1989). Since its inception, Monte Carlo methods have been applied to many different fields of study, including data analytics.

### 2.1.3   Use of Monte Carlo Methods with Game theory

The literature provides some evidence of Monte Carlo methods that have been applied within a game-theoretic context. The main two uses of Monte Carlo methods, within this context, are (1) to aid in finding an approximate solution and (2) explore outcomes of a game in complex and/or stochastic environments. The problem domains of the game considered include computer and wireless networks (Cao, Wei, Zhao, & Huang, 2012; Sun, Chang, Hu, & Wang, 2015); automated vehicle routing and UAVs (Shin, Jang, & Tsourdos, 2017; Zohdy & Rakha, 2012); manufacturing (Gharehshiran & Krishnamurthy, 2010); power-plants (Hariyanto, Nurdin, Haroen, & Machbub, 2009; Moshkin & Sauhats, 2016; Street, Lima, Freire, & Contreras, 2011); construction (Aliahmadi, Sadjadi, & Jafari-Eskandari, 2011); and robotics (Jang et al., 2018b). Monte Carlo methods have been applied with the context of non-cooperative, cooperative, and hedonic games, which we discuss in turn.

#### 2.1.3.1   Non-cooperative Games

Monte Carlo games have been used to find the solution of non-cooperative games, especially board games, since the rise of personnel computer and computing power (Bruce, 2008).

Monte Carlo tree searches have been applied to solving board games like chess and Go. Recently, the AlphaGo program, which used Monte Carlo trees and other Artificial Intelligence techniques, was the first program to beat a human world champion of Go (Silver et al., 2017). Though games like chess and go are thousands of years old (Shenk, 2007), there is no known solution to either game. Chess computer programs have been constructed since the 1950s (Shannon, 1950), but the sheer number of possibilities in either game makes it impossible to solve even with modern computing power. This phenomenon is referred to as the *curse of modeling* (Gosavi, 2003, p. p212). It is the curse of modeling that limits the number of players we consider within this research due to the sheer number of possible partition sets to consider; this is discussed in more detail in the method section.

### *2.1.3.2 Cooperative Games*

Monte Carlo methods have also been used in the cooperative environment for both finding the approximate solution and exploration of the game environment. Examples of using Monte Carlo methods to solve cooperative games include Fatima, Wooldridge, and Jennings (2002), Gharehshiran and Krishnamurthy (2010), and Moshkin and Sauhats (2016). Fatima et al. (2002), who used Monte Carlo simulation for computing the approximating the Shapley value in a voting game, used this approach to try and find a reasonable solution in polynomial time. Gharehshiran and Krishnamurthy (2010) used a Monte Carlo simulation to demonstrate the performance of an algorithm used to solve a non-super-additive cooperative game of a sensor network in a manufacturing plant. Finally, Moshkin and Sauhats (2016) used a Monte Carlo method to find the solution of a power-plant game.

Monte Carlos simulation has also been used to explore the outcomes of complex cooperative games (arguably, any game with more the three players could be considered complex). Hariyanto et al. (2009) use Monte Carlo simulation to simulate a game of electric power generation; the players are power plant owners who make plans on their electric power generation to maximize their playoff in a crowded, competitive marketplace. Aliahmadi et al. (2011) used Monte Carlos simulation to analyze the risk associated with a game of cooperation amongst those involved in a construction project. Street et al. (2011) also used Monte Carlo simulation to investigate the risk of an energy market game; their game focuses on the introduction of renewable energy sources into Brazil's energy market. Karmperis, Aravossis, Sotirchos, and Tatsiopoulos (2012) used a Monte Carlo simulation to investigate the outcomes of a stochastic revenue-cost-sharing cooperative game.

Hedonic games are a subset of NTU cooperative game theory. There have been several applications of Monte Carlo methods to hedonic games: Shin et al. (2017), Jang, Shin, and Tsourdos (2018a), Cao et al. (2012), and Cao and Wei (2012). In all these cases, the authors used the Nash stability criteria because, arguably, it is easier to solve than core stability.

Another approach related to Monte Carlo methods in a game theory context is algorithmic game theory or mechanism design. As its name suggests, it focuses on the design of a game, especially in the context of having complete control of the players' behavior, i.e., bots. For more information on Mechanism Design, see Dash, Jennings, and Parkes (2003).

Our application of Monte Carlo methods both solves and explores hedonic games. Also, we consider core stability, not Nash stability; hence, we believe that we provide a novel application of Monte Carlo methods to hedonic games.

## 3   Method

Our approach to finding the empirical properties of hedonic games is to generate millions of random games and solve them. Specifically, we are looking at hedonic games with strict preferences, and we use the core stability solution approach. To be able to do this, we first must be able to represent a hedonic game in a form in which it is easy to generate and solve a random game. We discuss the form of the game first and then how it is used in our Monte Carlo experiment.

### 3.1 Constructing a random hedonic game

The generation of a random hedonic game can be done by randomly generating a preference relationship for each of the n players. Our intention here is to show that any hedonic game, with strict preferences, can be represented at n x $2^{n-1}$ matrix, where each column is a permutation of the numbers $\{1, 2, …, 2^{n-1}\}$. Thus, if we generate a random matrix, with the permutation criteria, we have generated a random hedonic game, with strict preference. A quick note on notation, the grand coalition of all players is denoted by N = {1, 2, …, n} and n is assumed to be finite.

**Lemma 1**. Each player is a member of exactly half of all subsets.

**Proof**. Consider the set of all subsets that do not contain player 'i', $2^{N\setminus\{i\}} \subset 2^N$, and the set of subsets that do contain player "i," $2^{N\setminus\{i\}}(i) = \{S \cup \{i\} : S \subseteq N\setminus\{i\}\} \subseteq 2^N$, which is a slight abuse of the notation. By their definition, these two sets are disjoint, $2^{N\setminus\{i\}} \cap \left(2^{N\setminus\{i\}}(i)\right) = \emptyset$, and cover all possible subsets, $2^{N\setminus\{i\}} \cup \left(2^{N\setminus\{i\}}(i)\right) = 2^N$. There exists a bijection function between these two subsets of the form $f: 2^{N\setminus\{i\}} \to 2^{N\setminus\{i\}}(i)$ where $f(S) = S \cup \{i\}$ for all $S \in N\setminus\{i\}$. Since and $\left|2^{N\setminus\{i\}}\right| = \left|2^{N\setminus\{i\}}(i)\right|$ and they cover $2^N$ then $|2^N| / |2^{N\setminus\{i\}}(i)| = 2$. QED.

**Corollary 1.** Each player is a member of exactly $2^{n-1}$ subsets

Following the notation of Iehlé (2007), for each $S \subseteq N$ can be represented as the binary vector $\mathbf{1}^S \in \{0,1\}^n$, where the coordinates equal one if the corresponding player is a member of S and zero otherwise; that is, $\mathbf{1}^S_i = 1 \; if \; i \in S \; o/w \; 0$. Now every binary vector represents a binary number, where $i^{th}$ digit of a binary number is represented as $\mathbf{1}_i^S$. For example, the subset of three players that only contains players A and B can be represented as $(1, 1, 0)^T$, which in turn can be represented as 110, which is equivalent to decimal 6. Hence there exist a bijection between the subsets of N and the numbers {0, 1, …, $2^n$-1}. Note that zero represent the empty set since $\mathbf{1}^\emptyset = \mathbf{0}$. Also note the slight awkward ordering of the players because, traditionally, players are presented from left to right, whereas a binary number goes from right to left.

**Lemma 2.** The subsets that contain a player 'i' can be ordered from 1 to $2^{n-1}$; this can be represented by a bijection $h_i: 2^{N/\{i\}}(i) \rightarrow \{1, 2, \ldots, 2^{n-1}\}$ for finite $n \in \mathbb{N}$.

**Proof.** Consider the set $N\setminus\{i\}$, there exists a bijection between its subsets and {0, 1, …, $2^{n-1}$-1} from the discussion above. Since there is a bijection between $2^{N\setminus\{i\}}$ and $2^{N\setminus\{i\}}(i)$, there exists a bijection between $2^{N\setminus\{i\}}(i)$ and the set {0, 1, …, $2^{n-1}$-1}. Simply add one to this bijection to obtain the result. QED.

What lemma 2 implies is that an isomorphism exists between the subsets that contain a certain player and a finite list of numbers from 1 to $2^{n-1}$. Now we are going to show that for a vector $V \in \mathbb{N}^{2^{n-1}}$ can be constructed to represent the preferences of a player in a hedonic game.

**Lemma 3.** In a hedonic game with strict transitive preferences, there exists a numbering of the subsets that contain a certain player, say "i," defined by the function $V_i: 2^{N/\{i\}}(i) \rightarrow \mathbb{N}$, such that for that player's preferences:

$$S, T \in 2^{N\setminus\{i\}}(i) \text{ and } S \succ T \Leftrightarrow V_i(S) > V_i(T)$$

**Proof**. A finite set with strict transitive preferences implies the set is well-ordered (every subset has a least element based on the preference relation). Consider the least element of $2^{N/\{i\}}(i) \ni L_1$, i.e., $S \succcurlyeq L_1$. Define $V(L_1) = 1$. Consider the least element of $\left(2^{N\setminus\{i\}}(i)\right) \setminus L_1 \ni L_2$; define $V(L_2) = 2$. Repeat this process until all $2^{n-1}$ elements have a defined $V(.)$ value. This function holds for the conditions stated above. QED.

**Corollary 1.** The function V(.) is a bijection between $2^{N\setminus\{i\}}(i)$ and $\{1, 2, \ldots, 2^{n-1}\}$.

**Proof.** This follows because of strict preferences.

What the previous three lemmas allow us to do is create an isomorphism between the preference relation of a player and a vector $\in \mathbb{N}^{2^{n-1}}$. If all these vectors are joined they form a matrix $M \in M_{n \times 2^{n-1}}$.

To help understand this representation of a hedonic game as a matrix, let us consider a simple example of three players 'A', 'B,' and 'C.' Let us assume that the players' strict preferences are:

$$\{A\} \succ_A \{A, B\} \succ_A \{A, B, C\} \succ_A \{A, C\}$$

$$\{A, B, C\} \succ_B \{A, B\} \succ_B \{B, C\} \succ_B \{B\}$$

$$\{A, C\} \succ_C \{A, B, C\} \succ_C \{C\} \succ_C \{B, C\}$$

First, let us consider the preferences of the first player 'A.' The following table shows how the indexing generated for preference ordering for player 'A' and how the related subsets are indexed.

**Table 1.** Example preferences

| S | If $A \in S : T = S\setminus\{A\}$ | Binary representation of T | Decimal Equivalent of T | Add one | Preference Ordering of A |
|---|---|---|---|---|---|
| ∅ | - | - | - | - | - |
| {A} | ∅ | 00 | 0 | 1 | 4 |
| {B} | - | - | - | - | - |
| {A, B} | {B} | 10 | 2 | 3 | 3 |
| {C} | - | - | - | - | - |
| {A, C} | {C} | 01 | 1 | 2 | 1 |
| {B, C} | - | - | - | - | - |
| {A, B, C} | {B, C} | 11 | 3 | 4 | 2 |

This results in a vector $(4, 1, 3, 2)$, which represents the preference relation of player 'A' for the subset that it is a member. Again, we suffer from the slightly awkward ordering of the numbers because the player ordering goes from right-to-left, e.g., $\{A, B, C\}$, whereas binary digits go from left-to-right. The vectors, from each of the three players, can be combined to create a matrix:

$$\begin{pmatrix} 4 & 1 & 3 & 2 \\ 1 & 2 & 3 & 4 \\ 2 & 1 & 4 & 3 \end{pmatrix}$$

Where the players are the row index and the subsets, for which the player is a member, are indexed for the columns, and the players form the column index. Our results imply:

**Lemma 4.** A finite hedonic game with strict transitive preference can be represented as a matrix $\{a_{ij}\} = A \in M_{n \times 2^{n-1}}$ were "i" represents the players and "j" is the decimal represents that the subsets that contain "i:"

$$a_{ij} = V_i\left(h_i^{-1}(j)\right) \in \mathbb{N} \; s.t \; i \in \{1, 2, \ldots, n\}, j \in \{1, 2, \ldots, 2^{n-1}\}$$

$$a_{ij} \in \{1, 2, \ldots, 2^{n-1}\} \ s.t. \ \bigcup_{j \in \{1,2,\ldots,2^{n-1}\}} a_{ij} = \{1, 2, \ldots, 2^{n-1}\}$$

$$\forall i, j \ \forall k \in \{1, 2, \ldots, 2^{n-1}\} \ a_{ij} = a_{ik} \Rightarrow j = k$$

The proof follows from the discussion above. Note that since $h_i(.)$ is a bijection its inverse exists.

**Corollary 2.** Any matrix of the form given in Lemma 4 represents a finite hedonic game with strict preferences.

**Proof.** Here, we only reverse the bijections, h, for each player, to create a preference relationship such that:

$$V_i\left(h_i^{-1}(j)\right) >_i V_i\left(h_i^{-1}(k)\right) \Rightarrow h_i^{-1}(j) = S >_i T = h_i^{-1}(j) \ \forall j, k \in \{1, 2, \ldots, 2^{n-1}\}$$

Ours is not the only way to represent a hedonic game in matrix form. Karakaya (2011) points out that there is a normal-form representation of hedonic games. Since the highest number of coalitions in a coalition structure is 'n,' Karakaya's form allows each player to choose one of 'n' coalitions to join, so each player's strategies set is selecting which of the 'n' coalitions to join. The payoffs are determined by the resultant coalitions, i.e., for a given strategy profile, if player one and two are the only ones to choose coalition one, then their payoffs for that strategy profile are their equivalent preference value (discussed above). It assumed that players are not vetted before gaining membership to the coalition, i.e., they could freely choose which existing coalition they wish to join. Since there is an assumption of free movement, amongst existing coalitions, then the Nash stability criterion is appropriate for this game (the core stability concept does not work in this case). Constructing a random game with this normal-form version of a game would be difficult due to the need for consistency between strategy profiles outcomes, i.e., all strategy profiles that result in the same coalition must have the same payoffs for the members of that coalition. Thus, constructing this normal-form game is at least $O(n^n)$ in computational complexity.

The matrix form of a hedonic game is effectively a finite collection of permutations, which can be counted using the twelvefold way (Stanley, 1997). However, our concern here, with our Monte Carlo approach, is not the generation of all possibilities but only the generation of a predetermined number of random hedonic games.

### 3.1.1 Generating a Random matrix

If we generate a random matrix of the form given in lemma 4 then we have generated a random hedonic game with strict transitive preference. This can be achieved by randomly shuffling the numbers 1, 2, …., $2^{n-1}$ to create a vector, repeating this to create n vectors, then combining the transpose of vectors to form a matrix.

Considering that millions of games are generated using our Monte Carlo analysis approach, it is important to have a pseudo-random number generator that has a limited number of biases as possible. The current de facto standard used in Modeling and Simulation is the Mersenne Twister (Matsumoto & Nishimura, 1998), which we used in our Monte Carlo C++ program.

The Mersenne Twister has a period of $2^{19937}$. This period significantly exceeds requirements, for random numbers, which are estimated to be only $2^{40}$. The approach to generating a random matrix was as follows: generate a $2^{n-1}$ x n matrix filled with pseudo-random numbers that follow a standard uniform distribution; replace each random number with its ranking in the current row. In the unlikely event of two numbers in a row are exactly the same, the second is replaced with a new random number.

### 3.1.2 Connection to NTU games

An observant reader might notice that our matrix representation approach means that every coalition has an associated value vector associated with it. Thus, a characteristic function can be constructed that maps the subsets to a vector that has value for each agent. This efficiently creates a type of non-transferable utility (NTU) where the coalitions have a fixed payoff vector. For more details on the relationship between hedonic and NTU games, please see Iehlé (2007, p. p. 183). Since every hedonic game represented an NTU game, we see hedonic games as a subset of NTU games; this viewpoint is also expressed by others (Chalkiadakis et al., 2011).

### 3.1.3 Possible number of different hedonic game

Given that our Monte Carlo approach randomly samples hedonic games, it is reasonable to ask how many different possible hedonic games there are. Assuming that player labeling matters, then the number of possible hedonic games, with strict preferences, is:

$$\left((2^{n-1})!\right)^n$$

For two-players, this means that there are only four possible games; for three-player games, there are 13,824; for four-players games, $2.2 \times 10^{18}$; and for five-players games is $4.0 \times 10^{66}$. To put these numbers into perspective, consider a 10 Penta-Hertz computer processor, which is far more power than any computer currently in existence, that was able to generate a single game at each computational step, which would be impractically efficient, then it would take that computer processor $9.2 \times 10^{35}$ life-times of the universes (13 billion years) to completely determine all possible games of five-players. Determining all possibilities for greater than four players is simply an impossibility.

There are several ways of reasonability reducing this number. For example, Ballester (2004) points out that individual rational players will never join a coalition that is less preferred than their singleton set. As such, the ordering of the coalition ranked below the singleton set, for each player, can effectively be ignored. For example, consider game (a) in Table 2, player A considers the singleton set the most preferable, so it does not matter how they rank the other three possible coalitions that contain A. Each hedonic game can, thus, reduced to *Individually Rational Coalition lists (IRCL).*, as Ballester (2004) describes it. However, the number of possible hedonic games excessively large. If player ordering does not matter, then there are several equivalencies among the games, so the number of possibilities could be reduced further.

### 3.2   Finding the core set of a game

Determining if the core of a hedonic game is non-empty is NP-Complete (Ballester, 2004), under certain favorable circumstances. As a result, solving a game with only a score of agents can be computationally intractable (Hajduková, 2006). Given this difficulty in finding empirical results, the focus on the community has been on finding analytical results. These analytical results usually focus on finding necessary and sufficient conditions for the core existence. However, from an empirical viewpoint, these conditions can be just as hard to determine if a game satisfies them as if determining core set using a brute computational force; for example, the notion of pivotal balance, a necessary and sufficient condition for a non-empty core, requires the determining and checking of all I-balanced collections of coalition, which in turn, requires checking the balance of these coalition collections (Iehlé, 2007).

As already mentioned, the focus of this paper is the empirical study of hedonic games, and our approach is the generation of millions of hedonic games and their core partitions if they have one. This section discusses our approach to finding the core of a game. We use the term "solve" to mean finding the core partitions. There are several algorithms to solve hedonic games, usually requiring some limitations on its form, but all require exponential growth in computational time to solve larger games (Ballester, 2004). Our approach is the simple brute-force, or naïve, algorithm approach (Chalkiadakis et al., 2011). We generate a partition, then check it against every possible coalition to determine if it is blocked; if it is not blocked by any coalition, then the partition is part of the core. A process diagram, showing how the algorithm works, is given in Figure 1.

[INSERT FIGURE 1 ABOUT HERE]

The process diagram outlines the algorithm, which contains two loops for solving a single randomly generated hedonic game. The outer loop iterates over all possible coalition structures, and the inner loop iterate over all possible coalitions. For a given coalition structure, the inner loop will check to see if it is blocked by each coalition. Note that we are using the V(.) function notation to define preferences, as discussed earlier in the paper. Initially, all coalition structures are assumed to be in a core partition; if a coalition structure is shown to be blocked, it is removed from the core. Once all coalition structures have been tested for all potential coalition blocking, what remains is core.

Though called the naïve algorithm approach, it is far from simple. For example, how do we determine that you have checked every possible partition of a particular game? The number of possible partition, $|\Pi|$, is Bell's number (Bell, 1938). Bell's number is, according to Dobiński (1877), the following formula:

$$B_n = \frac{1}{e} \sum_{k=0}^{\infty} \frac{k^n}{k!}$$

Its asymptotic formula is not simple, but it is known that $B_n > O(ne^n)$. To give the reader some understanding of this number, consider a fifteen-agent game: there are 1,382,958,545 possible partitions of 32,768 possible coalitions resulting in about 45 trillion checks needed to be made per game to determine which partitions are in the core.

Given a large number of possibilities, both the partitions and subsets must be ordered so that they can be logically and sequentially checked. Since we have already shown a bijection between the subsets and the first $2^n$ numbers, this ordering of the possible coalitions is simple enough. Creating an ordered list of the partition is more complicated, and we used the setpart2 algorithm (Djokić, Miyakawa, Sekiguchi, Semba, & Stojmenović, 1989). This algorithm uses the fact that any partition of players can be represented by a single integer, leading with a one. For example, 12123 represents a game of five players where players 1 and 3 are in a coalition, {1,3}; players 2 and 4 are in a coalition, {2, 4}, and player 5 is in its singleton coalition, {5}. This representation of a coalition structure ensures it is both disjoint and covers the set of players.

The set of partitions is atomic, in terms of combinatorics, (Rossi, 2015), and, as such, it is feasible that a more mathematically elegant algorithm exist other that setpart2 one we used. However, our concern is practical computational efficiency and not elegance.

Even with all the mathematical "tricks" in play, the algorithm described is still computationally intensive for a large number of players. There have been attempts to reduce the computational complexity of finding the solution of cooperative games, but these approaches mainly involve restricting the form of the game to ensure it as certain properties, for example, induced subgraph games, see Deng and Papadimitriou (1994). One improvement to our algorithm that could have been the inclusion of individually Rational Coalition Lists (IRCL)

(Ballester, 2004). IRCL notes that any coalition that gives a player a payoff less then their singleton payoff is not individually rational and can be ignored. The use of IRCL would have reduced the number of subsets that needed to be checked in the inner loop of the algorithm.

Though we did not implement IRCL, we did check the singleton sets against the partition first before the main algorithm was complete. This was done because it is relatively simple to do and only involves "n" checks.

### 3.3 Monte Carlo Experiment

The previous section describes the algorithm approach to finding the core set of a random hedonic game. In this section, we discuss a Monte Carlo experiment to collect empirical data about hedonic games. The experiment involves finding the core for randomly generated hedonic games. From the data collected from this experiment, statistics about the core of the games were generated and discussed in the results section.

The only variable considered in the experiment was the number of players in the game. In our experiment, the number of players under consideration was 2 to 13. We limited the number of players to 13 due to computation limitations. We call the number of players, of a game, the size of the game. For each game size, we created one million randomly generated hedonic games (with strict preferences) using the matrix approach described above.

Using the generated data from these runs, various statistics were determined, including empirical distributions of the core size. The reason that a million was chosen for the number of games to be generated for each game size was that this number is so vast that it is very unlikely that the resultant empirical distribution does not follow the same shape as the population distribution. Thus, we feel confident that the results presented here are representative of the population's statistics and distributions.

A million data points far exceed the power of any influential statistical test on underlying population probability distributions for each game size. As a result, any differences between the empirical distribution and the hypothesis distribution will result in being significant when conducting hypothesis tests. There are likely to be differences due to the rounding errors that occur when generating the statistics; as a result, hypothesis testing becomes meaningless with one million data points. This is a common problem within the field of data analytics, and, as such, we avoid conducting hypothesis tests on our results.

A pilot study of this work conducted by Collins, Thomas, and Grigoryan (2019). This pilot study only considered a maximum size of seven, and only 50,000 runs were completed for each size. The pilot study did produce output distributions which followed the similar shape as those found in the paper; thus, the pilot study acts as a verification check of our data output.

Data was not collected from games of size one due to the triviality of the results, i.e., this only one possible coalition, so its corresponding coalition structure is automatically a core partition. We did complete runs for size two even though there is also always a core of size one; these runs acted as a verification check.

An overview of the complete experimentation process is shown in Figure 2. The figure shows three major tasks of the experiment. First, a random game (matrix) is generated, then its core found, and, finally, statistical analysis is conducted on the output. The program for generating the random games and finding their core was written in the C++ programming language to ensure the best memory management possible (of the current standard programming language). The statistical analysis was done using the R programing language using RStudio.

[INSERT FIGURE 2 ABOUT HERE]

The runs themselves were completed on three Dell Precision machine with Intel® Xeon® Gold 6146 CPU @ 3.20 GHz and 128 GB RAM. As mentioned earlier, there are computational limitations on the size of the game that can be considered. On the computer machine described here, we estimate it would take approximately 500 years to find the core of a single game of size 20. Though this game size might seem relatively low for such a large computational requirement, a reader should bear in mind that the game of chess, with only two decision-makers, 32 pieces, and 64 squares, has not solved due to its computational complexity (Shenk, 2007).

Since the core of about one million games were solve for each game size, various empirical distributions could be generated. Our focus is on the distribution of the size of the core, i.e., how many partitions were not blocked for a given game.

## 4   Result

Twelve million random hedonic games were generated using our approach for our empirical analysis. The games were grouped by the number of players in each game, and each group contained a million games. For each game, the core partition set was found. We refer to the size of the core partition set as the core size. The distribution of these core sizes is shown in Table 2. For completeness, the table includes the manual calculation of core size for games of one player.

**Table 2.** Empirical core size result for games of one to thirteen players

| Players | Core Size |
|---------|-----------|
|         |           |

|    | 0      | 1       | 2       | 3       | 4      | 5      | 6     | 7     | 8     | 9+ (max observed =18) |
|----|--------|---------|---------|---------|--------|--------|-------|-------|-------|------|
| 1  | -      | 1 M     | -       | -       | -      | -      | -     | -     | -     | -    |
| 2  | -      | 1 M     | -       | -       | -      | -      | -     | -     | -     | -    |
| 3  | 7,170  | 952,117 | 40,713  | -       | -      | -      | -     | -     | -     | -    |
| 4  | 19,566 | 878,328 | 96,731  | 5,253   | 121    | 1      | -     | -     | -     | -    |
| 5  | 33,593 | 798,176 | 150,855 | 16,193  | 1,115  | 65     | 2     | 1     | -     | -    |
| 6  | 46,235 | 722,188 | 194,521 | 32,228  | 4,235  | 520    | 63    | 9     | 1     | -    |
| 7  | 56,991 | 656,425 | 225,495 | 49,667  | 9,410  | 1,656  | 296   | 56    | 4     | -    |
| 8  | 64,401 | 599,949 | 246,387 | 67,854  | 16,532 | 3,786  | 862   | 183   | 38    | 8    |
| 9  | 68,351 | 553,564 | 259,432 | 85,142  | 24,574 | 6,546  | 1,780 | 441   | 133   | 37   |
| 10 | 70,890 | 514,170 | 266,404 | 100,528 | 33,469 | 10,180 | 3,045 | 903   | 267   | 144  |
| 11 | 72,094 | 483,108 | 270,061 | 112,414 | 41,453 | 13,973 | 4,628 | 1,480 | 529   | 260  |
| 12 | 71,102 | 455,585 | 270,138 | 123,530 | 49,954 | 18,914 | 6,937 | 2,483 | 873   | 484  |
| 13 | 69,682 | 432,851 | 269,424 | 132,877 | 57,712 | 23,021 | 8,916 | 3,343 | 1,317 | 857  |

Table 4. shows three phenomena relating to the core size and number of players in the game. Firstly, the number of games with an empty core increases from 7,170 in the three-player game to 69,682 in the thirteen-player game; however, this increase is not strictly monotone. Secondly, cores with size one also decrease, which we can see falls from 1 million in the one-player game to 432,851 in the thirteen-player game; this decrease is monotonic. Thirdly, the number of core, sized two and more grows dramatically, for instance, core sized two starts from 40,713 in the three-player game to 269,424 in thirteen-player game. The interpretation of these results is that, despite all concerns about the empty core of hedonic games in the research literature, we speculate that as the game size increases, the chance of having empty cores decreases. 4.1 Fitted distribution information

Using the empirical data, we wished to determine the best-fitted distributions to the core size. Fitting distribution is a process of finding a mathematical function in which represents the statistical variable (Ricci, 2005). Generally, three steps apply for this process; (1) find the distribution choices, (2) estimate the parameters, and (2) evaluate the quality of the fit. We used the R programming language, a powerful and flexible language used for statistical computation, to find fitted distribution to the data set through these three steps (Delignette-Muller, Dutang, Pouillot, & Denis, 2014; Venables & Ripley, 2013).

Before going through the fitting distribution process and evaluating distribution, it is generally acceptable to choose suitable distribution candidates for the whole process. Explanatory data analysis was use to explore the data and observe its empirical distribution. We represent a histogram for only games of 5, 9, 12, and 13 players in Figure 3. The graphs show the sprawling of the distributions as game size increase, this was expected.

[INSERT FIGURE 3 ABOUT HERE]

Histogram for each game seems similar to the lognormal distribution, which was our initial hypothesis. To be more accurate about our distribution candidate, it is recommended to use descriptive statistics beyond just empirical plots. The two factors that can provide more information about distribution are skewness and kurtosis. Kurtosis value is a measure of the weight of the tail, and Skewness value gives information about symmetry in the empirical distribution. The normal distribution has a skewness value of zero and a kurtosis value of three. One suitable tool for showing skewness-kurtosis relation in empirical data is Cullen and

Frey graph (Cullen & Frey, 1999). Cullen and Frey's graph shows us potential distribution candidates for our data set. On this plot, common continuous distributions like normal, uniform, exponential, logistic, beta, lognormal, gamma, and Weibull are represented. The observation from each data set marked by a single holo circle on the plot.

[INSERT FIGURE 4 ABOUT HERE]

By looking at Figure 4, it is hard to tell which observations belong to which game size; however, there is a pattern. The smallest game size considered was three, its observation is in the bottom-right corner, with a kurtosis of nineteen. For the next three game sizes both the positive skewness value and kurtosis reduced. The results seem to cluster around a kurtosis value of approximately six (skewness of two) for the remaining games from size six to 13 players. This means that the distribution of core sizes seems to be converging to one of three common right-skewed distributions of lognormal, gamma, and Weibull distributions. There is a lack of symmetry in the distributions because of the non-zero skewness value. The kurtosis value for most games are far from the normal distribution, and it reveals that none of them follow a normal distribution.

Though not directly clear from the graph, the game sizes seem to hit a minimum around a size of eight, for both kurtosis and skewness. As the game sizes increase, post this minimum, they seems to be moving towards the gamma line and, potentially, the exponential point. We are attempting to run larger game sizes to confirm this hypothesis. This can be shown more clearly using cumulative distribution function (CDF) plots, shown in Figure 5.

[INSERT FIGURE 5 ABOUT HERE]

So, based on the CDF plot, we would suggest that the lognormal distribution is not fitted to the center or even tail of our empirical distributions. Gamma and Weibull distribution can be most likely fit the right tail but not good at the center of data. However, this bad fit of the center is likely due to the discrete nature of our dataset outputs (i.e., the core partition size is always a whole number); we believe that as the game size increases the effects from the data's discrete nature will be reduced making a better fit to the two distributions.

### 4.1.1 Goodness-of-fit statistics

The third step of understanding fitted distribution to the empirical data is to understand how far the distances between empirical distributions and fitted parametric distributions are. We use several goodness-of-fit statistics: Kolmogorov-Smirnov, Cramer-von mises, Anderson-darling statistics, Akaike Information Criterion (AIC) and Bayesian Information Criterion (BIC). Detailed information about these items can be found in D'Agostino (1986), Akaike (1974) and, Schwarz (1978).

The Anderson-darling statistic (minimum distance estimation) is the best in situations where both tails and bodies of distribution equally important for us. It is mostly used in a situation when we want to test a family of distributions. Kolmogorov-Smirnov and Cramer-von Mises statistic do not considering the complexity of the model when comparing distributions so they are practical when the number of parameters is equal in all distributions. AIC calculates the quality of the estimated model by considering the amount of information loss in a model. So, the preferred model between a set of models is the model with minimum AIC value. BIC is also used for model selection, where there is a finite number of models with a model with a lower BIC being preferred model. Researchers believed that AIC is a better criterion for model selection than BIC under certain assumptions (Burnham & Anderson, 2004; Vrieze, 2012; Yang, 2005). We provide goodness-of-fit information about Weibull and Gamma distributions in Table 3; again, we only provide a sampling of the results for five, nine, 12, and 13 players. In all cases, we are looking for the lowest value.

**Table 3.** Goodness-of-fit statistics and criteria calculation for Weibull, and Gamma distributions in games of five, nine, twelve, and thirteen players

| Players | 5 | | 9 | | 12 | | 13 | |
|---|---|---|---|---|---|---|---|---|
| Distributions | Weibull | Gamma | Weibull | Gamma | Weibull | Gamma | Weibull | Gamma |
| Goodness-of-fit statistics | | | | | | | | |
| Kolmogorov-Smirnov statistic | 0.47 | 0.51 | 0.43 | 0.47 | 0.41 | 0.45 | 0.38 | 0.42 |
| Cramer-von Mises statistic | 47,640 | 54,5099 | 29,375 | 37,554 | 27,265 | 35,594 | 9,581 | 12,880 |
| Anderson-Darling statistic | 229,100 | 256,402 | 167,770 | 188,006 | 162,658 | 180,391 | 59,431 | 67,208 |
| Goodness-of-fit criteria | | | | | | | | |
| Akaike's Information Criterion | 1,987,804 | 2,227,955 | 2,774,379 | 2,685,493 | 3,030,436 | 2,834,909 | 1,403,075 | 1,350,543 |
| Bayesian Information Criterion | 1,987,827 | 2,227,978 | 2,774,403 | 2,685,517 | 3,030,460 | 2,834,933 | 1,403,097 | 1,350,565 |

All goodness-of-fit statistics indicate that the Weibull is the better fit; this was true for all game sizes. However, the goodness-of-fit criteria have mixed results. For smaller game sizes, both criteria indicate that the Weibull is a better fit; however, this flips to the Gamma being a better fit around eight players. Thus, making conclusions about the best choice between these two-parameter distributions becomes challenging and no conclusion are given here. However, in both distribution cases, the actual fitted parameters are changing with game size, as shown in Table 4.

Table 4. Weibull and Gamma distribution parameters for games of three to thirteen players

| Player | 3 | 4 | 5 | 6 | 7 | 8 | 9 | 10 | 11 | 12 | 13 |
|---|---|---|---|---|---|---|---|---|---|---|---|
| Weibull Parameters | | | | | | | | | | | |
| Shape | 3.52 | 2.28 | 1.67 | 1.34 | 1.14 | 1.02 | 0.96 | 0.82 | 0.8 | 0.79 | 0.89 |
| Scale | 1.1 | 1.16 | 1.21 | 1.27 | 1.33 | 1.39 | 1.46 | 1.5 | 1.56 | 1.64 | 1.76 |
| Gamma Parameters | | | | | | | | | | | |
| Shape | 5.88 | 2.26 | 1.37 | 1.02 | 0.84 | 0.75 | 0.7 | 0.58 | 0.57 | 0.57 | 0.66 |
| Scale | 5.69 | 2.08 | 1.19 | 0.83 | 0.64 | 0.54 | 0.48 | 0.37 | 0.34 | 0.33 | 0.36 |

Initially, the change in the parameters seems monotonic with a change in the game size; the shape parameters are decreasing, the Gamma scale parameter is decreasing, and the Weibull scale parameter is increasing. However, there seems to be a local minimum reach around twelve players with a reversal of direction for the Weibull shape, the Gamma shape and the Gamma scale. This reversal seems to coincide with the reversal found in the Cullen and Frey graph. When the shape value goes close to one then, for both Gamma and Weibull distributions, the shape of distributions will be close to the exponential distribution (Forbes, Evans, Hastings, & Peacock, 2011; Gupta & Kundu, 2001). As such, we hypothesis that the empirical distribution is moving towards an exponential distribution. Further research will be required to investigate this hypothesis.

Other Future works could include the analysis of other statistics from our data. For example, the distribution of average coalition sizes or average player payoff.

## 5  Conclusions

This paper outlines an approach for using Monte Carlo methods, a numerical approach, to find the core of millions of hedonic games. We use the term core to mean the set of core partitions of a given hedonic game. Our approach includes a method for generating a random hedonic game with strict preferences. The Monte Carlo experiment was intended to provide insight into hedonic games that are currently not achieved by analytical means. This paper outlines all the elements required to conduct a Monte Carlo experiment as well as discussing some of the results.

The results from the Monte Carlo experiment focus on the core size and, by obvious extension, core existence. The input variable for these experiments is the number of players in a given hedonic game, which we call game size. Solving the games numerically, as opposed to analytically, allowed for millions of different games to be solved. Our results show that: there is a monotonic decrease of games with a core of size one when you increase the game size; there is a monotonic increase in games with a core sized more than two; and a not strictly monotonic increase in games with an empty core or a core of size two. The statistical methods considered show that there is a high tendency for the core distribution of games of six players and more to follow a Weibull or gamma distribution.

Unlike the analytical insight provided by the majority of papers on hedonic games, this numerical insight gives us an understanding of what might be expected when dealing with actual hedonic games. As Aziz and Savani (2016) say, "An important area of future research is to model and capture realistic scenarios via hedonic games. There is a need to bring together the work on behavioral game theory and mathematical game theory." We believe that the approach outlined in this paper is a step forward towards this goal.

A real-world hedonic game may not have a prescribed structure and, more importantly, will need to incorporate human behavior models which will create games rich with a complexity that cannot be solved analytically, using our current mathematically understanding. As such, an empirical method must be employed to gain an understanding of such games.

# FIGURES

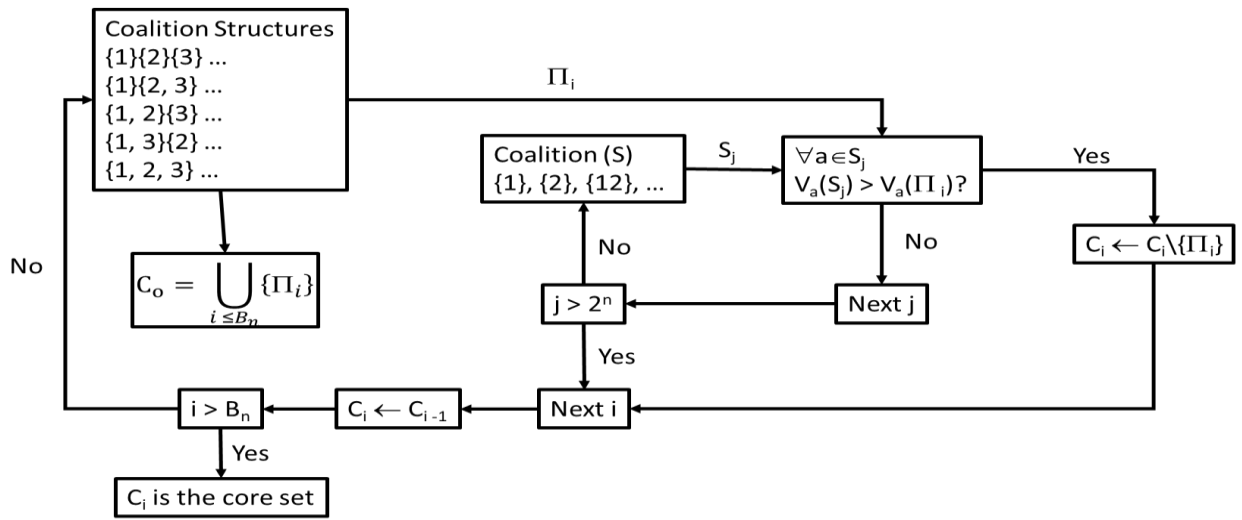

**Figure 1**: Process diagram of brute-force algorithm for determining the core of a random hedonic game

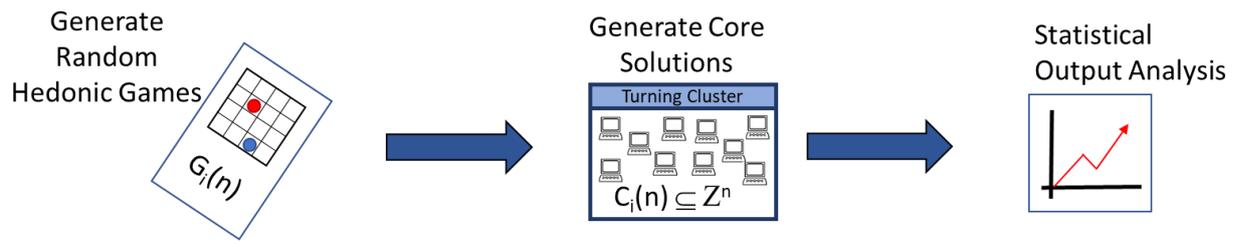

**Figure 2:** Overview of the experimental process

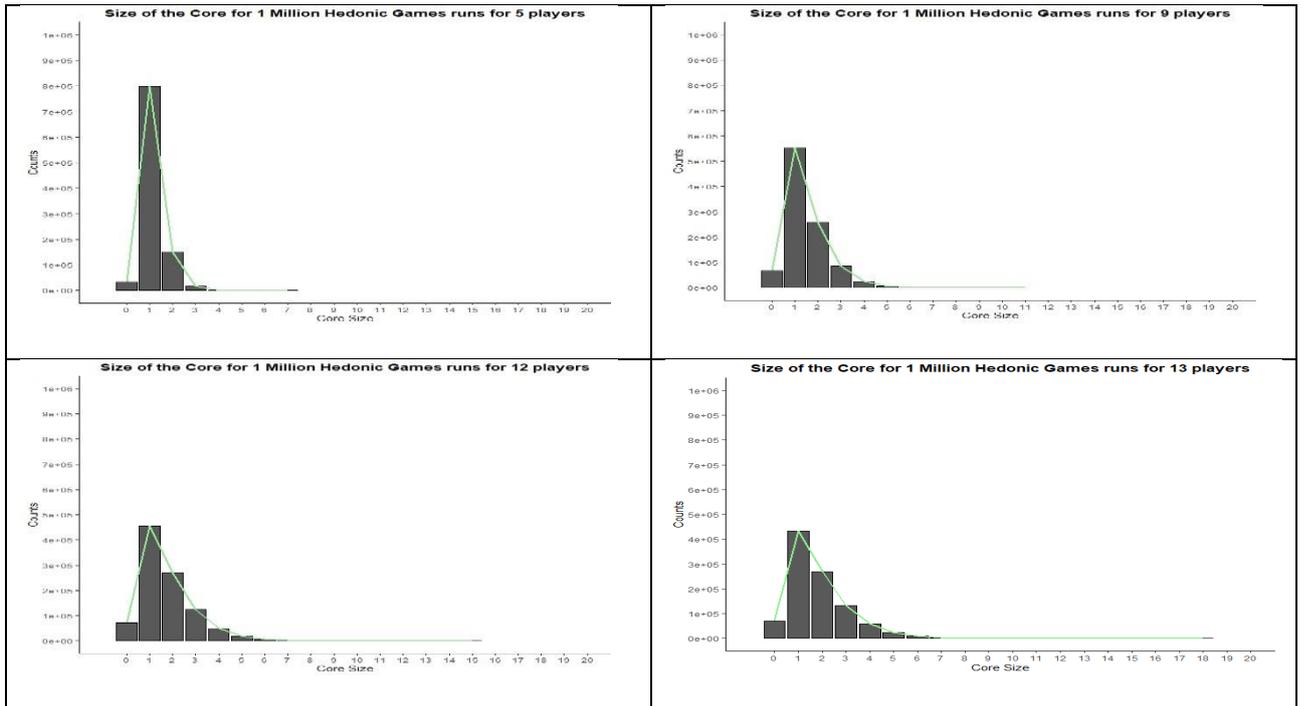

**Figure 3:** Distribution of core size for games of 5, 9, 12, and 13 players

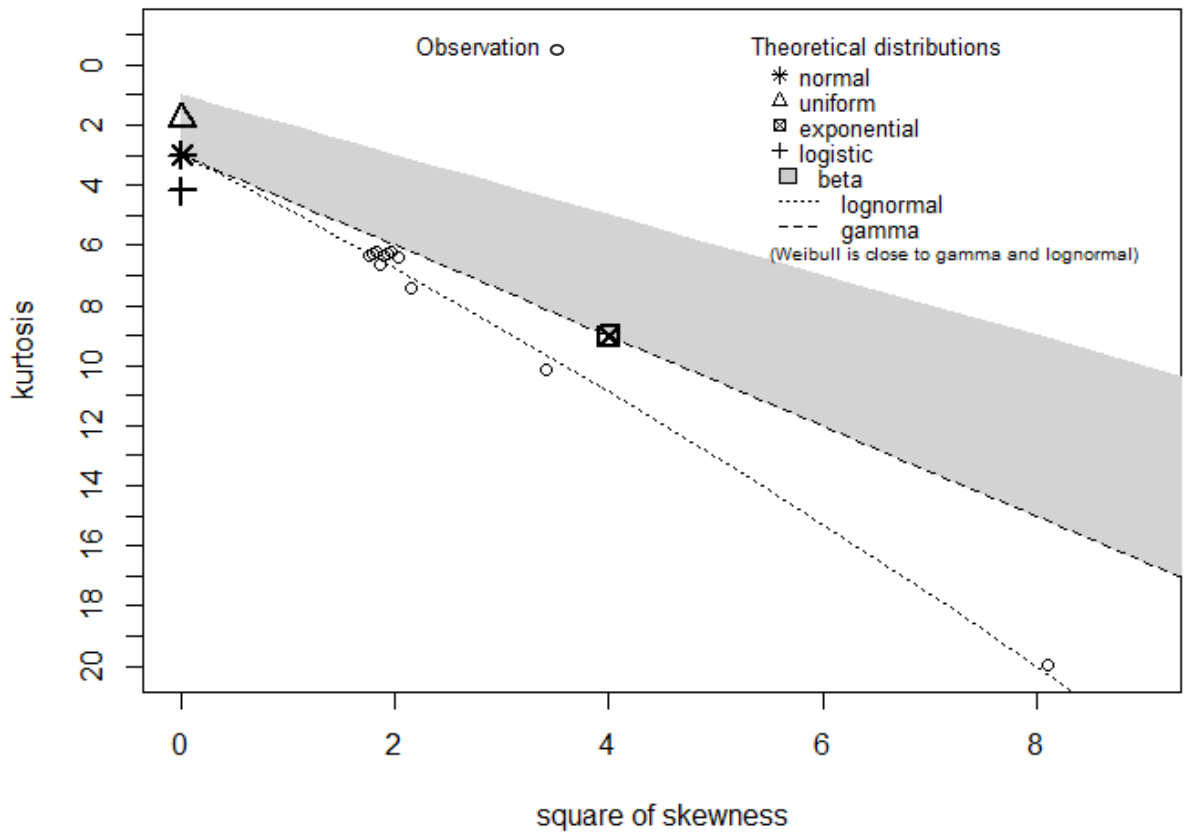

Figure 4: Cullen and Frey graph for games of three to thirteen players

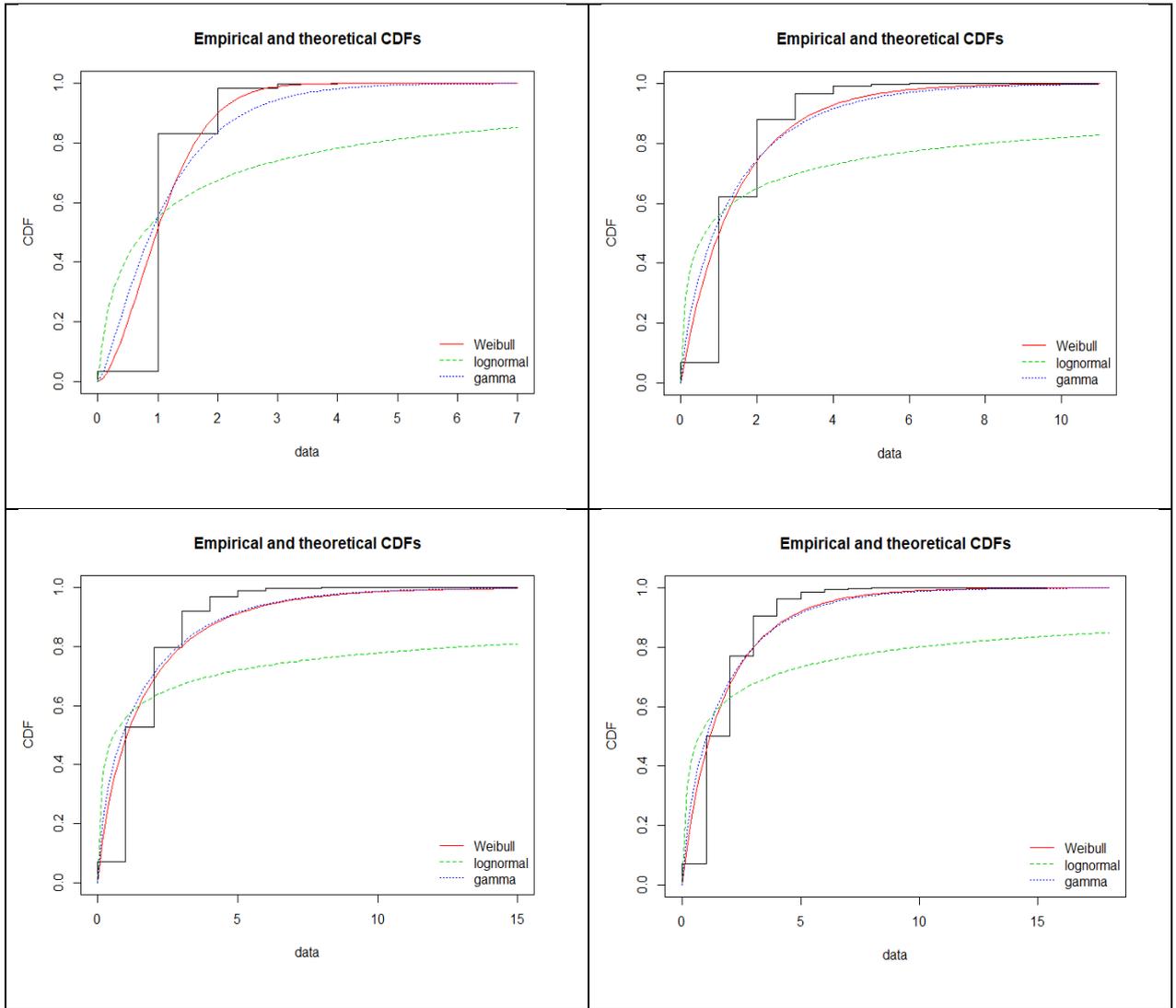

**Figure 5:** CDF plots for games of 5, 9, 12, and 13 agents from left-top to right-bottom